\UseRawInputEncoding
\documentclass[twocolumn,           % Format : preprint, twocolumn
               showpacs,            % Pacs : showpacs, noshowpacs
               nopreprintnumbers,     % Preprint: preprintnumbers,
               aps,                 % Society: ...
               prd,          	    % Journal Style : pra, prb, prc, prd, pre,
               letterpaper,             % Size : a4paper, ...
              groupedaddress,      % Affiliation (Title) : groupedaddress,
               nofootinbib,         % Footnote: footinbib, nofootinbib
               tightenlines,        % Remove additional spaces in a line
               floats,floatfix,      % Floating pictures and tables
               showkeys
               ]{revtex4-1}
               
\usepackage[toc,page]{appendix}
\usepackage{graphicx}% Include figure files
\usepackage{dcolumn}% Align table columns on decimal point
\usepackage{bm}% bold math
\usepackage{amsmath}
\usepackage{amsfonts,amssymb}
\usepackage{soul}
\usepackage{color}
\definecolor{v}{rgb}{0.6, 0.2, 0.8} %comentarios VM
\definecolor{MAGA}{rgb}{0.1, 0.43, 0.75}
\definecolor{jm}{rgb}{0.13, 0.48, 0.64}

\usepackage{xcolor}
\usepackage{enumerate}
\usepackage{float}
\usepackage{subfigure}
\usepackage{multirow,tabularx, booktabs}

\usepackage{orcidlink}

\usepackage{silence}
\begin{document}

\title{Constraining a model supported in Moir\'e gravity with a recent data release 
}

\newcommand{\orcidauthorA}{0000-0003-0405-9344} % Alberto
\newcommand{\orcidauthorB}{0000-0002-6356-8870} % Miguel
\newcommand{\orcidauthorC}{0000-0001-9310-2935}
\newcommand{\orcidauthorE}{0000-0002-8025-2763}

\author{J. A. Astorga-Moreno$^1$\orcidlink{\orcidauthorC}}
\email{jesus.astorga@cinvestav.mx }

\author{Miguel A. Garc\'ia-Aspeitia$^2$\orcidlink{\orcidauthorB}}
\email{angel.garcia@ibero.mx}

\author{A.  Hern\'andez-Almada$^3$\orcidlink{\orcidauthorA}}
\email{ahalmada@uaq.mx}

\author{E. A. Mena-Barboza$^4$\orcidlink{\orcidauthorE}}
\email{eri.mena@academicos.udg.mx}

\affiliation{$^1$Departamento de F\'isica, Centro de Investigaci\'on y Estudios Avanzados del IPN, Apartado Postal 14-740, 07000, CDMX, Mexico}

\affiliation{$^2$ Depto. de F\'isica y Matem\'aticas, Universidad Iberoamericana Ciudad de M\'exico, Prolongaci\'on Paseo \\ de la Reforma 880, M\'exico D. F. 01219, M\'exico}

\affiliation{$^3$ Facultad de Ingenier\'ia, Universidad Aut\'onoma de
Quer\'etaro, Centro Universitario Cerro de las Campanas, 76010, Santiago de 
Quer\'etaro, M\'exico}

\affiliation{$^4$ Centro Universitario de la Ci\'enega, Universidad de Guadalajara, Ave. Universidad 1115, C.P. 47820, Ed. de Investigaci\'on y Tutor\'ias, Ocotl\'an, Jalisco, M\'exico}

%-------------------------------------------------------------------------------------------------
%-------------------------------------------------------------------------------------------------
\begin{abstract}
Moir\'e theory is a revolutionary alternative arising from solid-state physics that can be extended to gravity, forming a framework called Moir\'e gravity. The idea behind it is that Einsteinian gravity has a validity limit and must be extended to Moir\'e gravity, in which the Friedmann equation changes and a new scalar field emerges. In this vein, in this paper, we propose an analytic solution for the Moir\'e scalar field coupled with the modified Friedmann equation, studying the viability and constraining the value for the main parameter of the scalar field and cosmology in general, through the use of   datasets that contain cosmic chronometers, type Ia supernovae, intermediate-luminosity quasars, and hydrogen II galaxies. Our results allow for an accelerated phase with two peaks, but also predict an extremely decelerated stage, with $q(z=0)=15.376^{+6.403}_{-4.692}$ when the characteristic parameter of the theory is $\jmath=0.677^{+0.001}_{-0.002}$, in contrast to other models that decelerate at $z=0$.  
\end{abstract} 
%\draft
\pacs{Dark energy, Cosmology, Moir\'e Gravity.}
%\date{=day}
\maketitle

%%%%%%%%%%%%%%%%%%%%%%%%%%%%%%%%%%%%
\section{Introduction}
%%%%%%%%%%%%%%%%%%%%%%%%%%%%%%%%%%%%

The exploration of the cosmos based on General Relativity (GR) led to the development of the Standard Cosmological Model ($\Lambda$CDM), which has impressed with observational evidence \cite{Perlmutter:1999,riee}. Actually, one of the greatest challenges for the $\Lambda$CDM model and in general modern cosmology, is a concise understanding of the Dark Matter (DM) and Dark Energy (DE), which phenomenologically is viewed as a perfect fluid with a time varying equation of state (EoS), associated in the $\Lambda$CDM model with the Cosmological Constant ($\Lambda$). Roughly speaking, both exotic components account for approximately 95\% of the Universe, with DE closely related to the current accelerated expansion observed at late times. 
Despite the many achievements of General Relativity (GR), problems in the cosmological standard model, such as $\Lambda$, $H_0$ tension, and Cosmic Microwave Background (CMB) anisotropy anomalies, have motivated the community to formulate alternative theories of gravity to address them from a new angle without distorting what GR has offered (see \cite{CosmoVerseNetwork:2025alb} for a compilation). One of these outstanding proposals is the Unimodular Gravity (UG) \cite{uni}, motivated by the appearance of $\Lambda$ as a constant of integration in the equations of motion allowing its value to be fixed as an initial condition; this may avoid the question why quantum fluctuations do not set the cosmological constant to the value $m^2_p$, since the estimation of the energy density for $\Lambda$ differs 120 order of magnitude compared with the one already observed \cite{Alcantara-Perez:2023jbv}. If we consider the physical motivations related to the possibility of a more realistic representation of the gravitational fields near curvature singularities and to create some first-order approximation for the quantum theory of gravitational fields, we have the $f(R)$ theories \cite{staro}, where the generalization of the Einstein-Hilbert (EH) Lagrangian density exhibits second-order curvature invariants. Another alternative high order gravity theory is Gauss-Bonnet gravity \cite{noj}, which couples a scalar field (inspired by String/M-theory) not only with the scalar curvature, but also with higher order curvature invariants, being relevant to explain non-singular early time cosmologies. 
In recent years, this kind of theory has experienced a renaissance in an attempt to explain the well-known enigma of the late-time accelerated expansion of the Universe, and, for $f(R)$ gravity, reflects consistency with the Planck-2018 data \cite{agaa}. 

In this direction, Moir\'e gravity \cite{moi1,1mg} shows potential in cosmology where the derived toy model, inspired in the analysis of a twisted and strained bilayer graphene that exhibits the Moir\'e geometric effects, which are quasi-periodic patterns \cite{MP,ray} appearing by interference when two curved identical patterns are overlapped in the presence of a small displacement. 

Graphene and other Dirac materials constitute a valuable condensed-matter framework for investigating emergent relativistic and geometric effects \cite{new1}. In the continuum approximation, the low-energy dynamics of graphene can be effectively described in terms of Dirac fermions, whose behavior is modified by lattice distortions that give rise to effective gauge fields and geometric degrees of freedom. In this context, strain and curvature can be interpreted as sources of effective gauge and geometric fields acting on the Dirac quasiparticles \cite{new2}. Furthermore, topological lattice defects play a significant role in the emergent geometric description: dislocations can be associated with effective torsion, whereas disclinations produce curvature. These results have contributed to the development of connections between the physics of Dirac materials and theoretical scenarios involving emergent gravitational phenomena \cite{new3}.

Topological structures and Moir\'e patterns offer promising condensed-matter platforms for exploring how geometric and gravitational-like phenomena can emerge from microscopic quantum degrees of freedom. In graphene, lattice distortions and topological defects can induce effective gauge fields, curvature, and torsion that govern the dynamics of Dirac quasiparticles. Similarly, twisted and strained bilayer systems can give rise to effective curved geometries and Moir\'e-induced gravitational behavior \cite{new4}. From the perspective of the Xons framework, such emergent phenomena can be regarded as macroscopic manifestations of an underlying quantum dynamics, in which matter and spacetime emerge collectively from a common set of fundamental degrees of freedom \cite{new0}.

We remark, despite the extensive literature, that Moir\'e patterns can not only be found in the treatment of materials and their properties, but there are research fields in medicine \cite{med}, crystallography \cite{crys}, and optics \cite{opt}, closely related to this phenomenon, adding to the many relevant contributions to what has been done in this scenario. 

Now, the Moir\'e physics in gravitational systems and hence in cosmology works considering two curved, similar layers of graphene as a metric system or space-times, which by definition guaranty the appearance of Moir\'e geometry, extensively studied in the context of a bilayer, together with the total action given by the sum of the action in each layer and the inter-layer one. Then, treating it with the EH action, we get two copies of the classical gravitational theory involving two different metrics (a bi-world emulating a $3+1$ dimensional manifold), making it possible to introduce a purely geometric inter-universe coupling term. Additionally, \cite{moi1} describes how this framework offers an effective cosmological constant that can be made arbitrarily small, constraining the two metrics involved via a scaling diffeomorphism, introducing the conformal relation through a scalar function $\phi$ known as the Moir\'e field.
We finally state that the physics outlined above is similar to the Moir\'e effect in condensed matter; as a consequence, two superimposed similar structures give rise to a superstructure with lower emergent energy scales than in the native theories. 

This manuscript is outlined with the following scheme: The section \ref{sec:cosmology} shows the mathematical details associated with the cosmology in a bi-world scenario, meanwhile the discussion about the treatment of observational samples for constraint the resulting free parameters in the model is given in Sec. \ref{sec:constraints}. Finally, our results and the respective final comments from the exposed in the paper are shown in Sections \ref{Results} and \ref{SD}, respectively. We henceforth use units in which $c=\hbar=k_B=1$.

%%%%%%%%%%%%%%%%%%%%%%%%%%%%%%%%%%%%
\section{Cosmology and Moir\'e Gravity} \label{sec:cosmology}
\begin{figure*}
   \centering
   \includegraphics[width=0.5\textwidth]{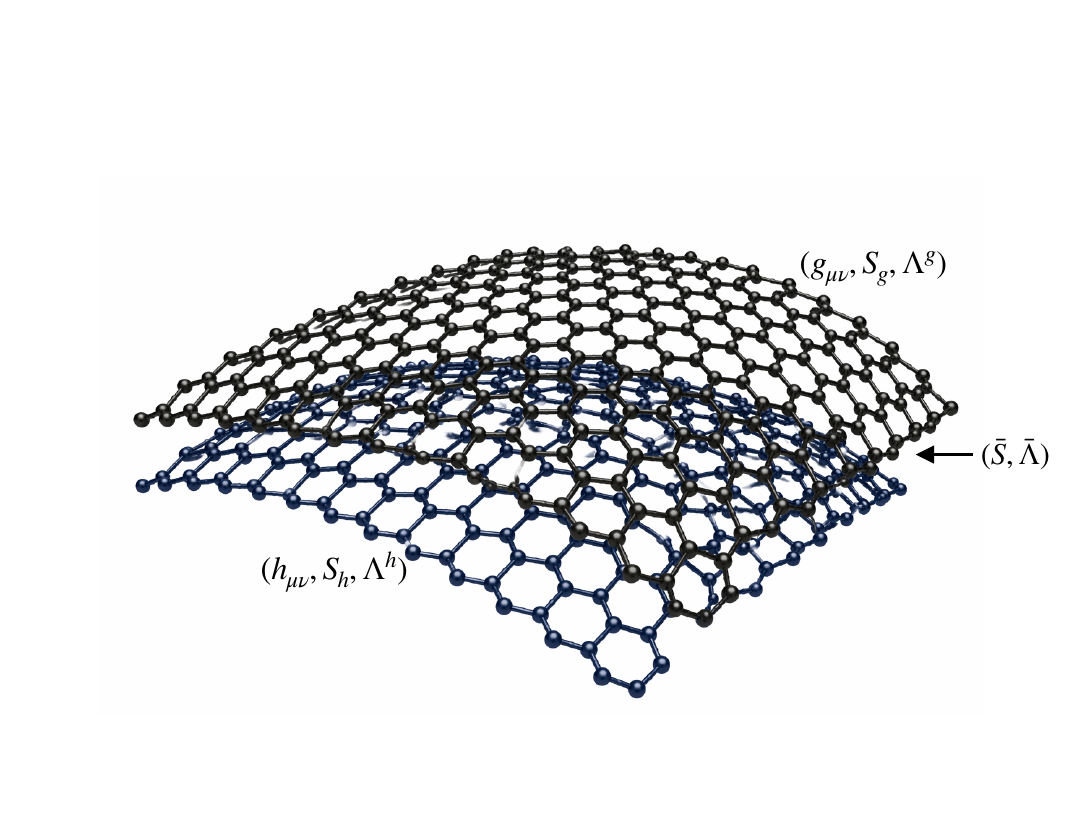}
   \caption{A twisted and strained bilayer graphene where the respective triad indicates the metric, action, and $\Lambda$ in each layer, and the ordered pair gives the action inter-layers and the $\Lambda$ associated. Also, we can imagine a curved layer of graphene as a surface covered by a hexagonal lattice of graphene.}
   \label{layers}
\end{figure*}

%%%%%%%%%%%%%%%%%%%%%%%%%%%%%%%%%%%%%%%%%%%%
The equations of motion (see Ref. \cite{1mg} for details), which allow us to study the dynamics in a bi-world context, are derived considering an action with two copies of the EH action for the metrics $h_{\mu\nu}, g_{\mu\nu}$ and an inter-universe element (see Figure \ref{layers} for details). 
Consequently, an effective form for the action comes with the substitution $h_{\mu\nu}\mapsto\phi^2 g_{\mu\nu}$ (meaning to be conformally related), in the presence of the dimensionless variable $\phi(t)=\tilde \phi( t)H^{-2} 
_0 $, bearing in mind that $\tilde \phi$ is the physical field Moir\'e, together with an adequate ansatz for $g_{\mu\nu}$, that allows us to decouple $\phi$ from the Ricci scalar. Thus, the action takes the form
\begin{eqnarray}
    &&S[g,\phi]=\int d^4\sqrt{\vert g\vert}\Big[R-6\frac{(1-\jmath)\phi^2-\jmath}{(1+\phi^2)^2}\partial_{\mu}\phi\partial^{\mu}\phi\nonumber\\&&-2\frac{\Lambda^{\eta}\phi^4+\Lambda^g+2\bar{\Lambda}\phi^2}{(1+\phi^2)^2}\Big].
\end{eqnarray}
Then, introducing a flat Friedmann-Lemaitre-Robertson-Walker (FLRW) universe, with the well-known metric $g_{\mu\nu}=(-1,a^2(t),a^2(t),a^2(t))$, we are in a position to reveal the modified Friedmann equations
\begin{eqnarray}
    3H^2=&& 8\pi G \Sigma_i \rho_i + 3\frac{(1-\jmath)\phi^2-\jmath} 
         {(1+\phi^2)^2}\dot{\phi}^2\nonumber\\
         &&+\frac{\Lambda^h\phi^4+\Lambda^g+2\bar{\Lambda}\phi^2}{(1+\phi^2)^2}, \label{f1}
         \end{eqnarray}
\begin{eqnarray}
    2\dot{H}+3H^2=&&-\left(8\pi G\right) \Sigma_i\omega_i \rho_i-3 \frac{(1-\jmath)\phi^2-\jmath}{(1+\phi^2)^2}\dot{\phi}^2\nonumber\\    
&&+\frac{\Lambda^h\phi^4+\Lambda^g+2\bar{\Lambda}\phi^2}{3(1+\phi^2)^2}, \label{f2}
\end{eqnarray}
where $H\equiv\dot{a}/a$, $G$ is the Newtonian gravitational constant and $\jmath$ is an appropriate constant.
We observe that $\Lambda^{g,h}$, $\bar{\Lambda}$ are the $\Lambda$'s in the universe with the respective metric and the inter-world $\Lambda$, respectively. Also, eqs. \eqref{f1}, \eqref{f2} are minimally coupled with the following Klein-Gordon-type equation for the Moir\'e field
\begin{eqnarray}
\ddot{\phi}=&&-3H\dot{\phi}+
\frac{\phi}{(1-\jmath)\phi^2-\jmath}\Bigg[\frac{(1-\jmath)\phi^2-\jmath-1}{1+\phi^2}\dot{\phi}^2 \nonumber\\
&&-\frac{2(\Lambda^h-\bar{\Lambda})\phi^2-2(\Lambda^g-\bar{\Lambda})}{3(1+\phi^2)}\Bigg].\label{Fi}
\end{eqnarray}
According to the Friedmann equations \eqref{f1},\eqref{f2}, the continuity equation reads 
\begin{equation}
    \dot{\rho}_i + 3H \rho_i (1+w_i)=0,
\end{equation}
where $\rho_i$ are the densities of the usual cosmological fluids and $w_i$ is the corresponding parameter in the EoS. Also, for the %Moir\'e field 
variable $\phi(t)$ %(see Appendix \ref{apendice})
\begin{eqnarray}
3[(1-\jmath)\phi^2-\jmath](\dot{\phi}H+\ddot{\phi})+3\phi\dot{\phi}^2(1-\jmath)+\nonumber\\
2\phi(\Lambda^h\phi^2+\bar{\Lambda})-
6\phi\frac{[(1-\jmath)\phi^2-\jmath]}{1+\phi^2}\dot{\phi}^2-\nonumber\\
2\phi\frac{(\Lambda^h\phi^4+\Lambda^g+2\bar{\Lambda}\phi^2)}{1+\phi^2}&=0,
\end{eqnarray}
with the aid of \eqref{Fi}, we have
\begin{align}
\frac{d\phi}{dz}&=-\frac{[(1-\jmath)\phi^2-\jmath](1+\phi^2)}{(1-\jmath)(1+z)\phi^2},\label{23}
\end{align}
Solving \eqref{23}, we get the analytical expression
\begin{eqnarray}
\mathcal{H}_1(\phi)=\frac{\sqrt{\jmath}}{2\sqrt{1-\jmath}}\ln{\Bigg[1+\frac{2\frac{\sqrt{\jmath}}{\sqrt{1-\jmath}}}{\phi-\frac{\sqrt{\jmath}}{\sqrt{1-\jmath}}}\Bigg]}-\text{arctan}[\phi]=\nonumber\\
\frac{\ln{(z+1)}}{1-\jmath},\label{jj}
\end{eqnarray}
also, if we consider the range $-\frac{\sqrt{\jmath}}{\sqrt{1-\jmath}}<\phi\leq \frac{3\sqrt{\jmath}}{\sqrt{1-\jmath}}$ together with $0<\jmath<1$, then the logarithm in $\mathcal{H}_1$ to first order, makes \eqref{jj} reads
\begin{eqnarray}
\mathcal{H}_2(\phi)=\frac{\jmath}{(1-\jmath)\cdot\Big[\phi-\frac{\sqrt{\jmath}}{\sqrt{1-\jmath}}\Big]}-\frac{\pi}{2}=
\frac{\ln{(z+1)}}{1-\jmath},
\end{eqnarray}
and\footnote{ For $\vert\phi\vert \gg1$, we have $\mathcal{H}_1(\phi)=\mathcal{H}_2(\phi)+\mathcal{O}(\epsilon)$, with $\epsilon \downarrow0$ and $\mathcal{H}_1(\phi)=\mathcal{H}_2(\phi)+\mathcal{O}(1)$, for $\vert\phi\vert<1$.}
\begin{equation}
\mathcal{H}_1(\phi)=\mathcal{H}_2(\phi)+\mathcal{O}\Bigg(\frac{2\frac{\sqrt{\jmath}}{\sqrt{1-\jmath}}}{\phi-\frac{\sqrt{\jmath}}{\sqrt{1-\jmath}}}\Bigg),
\end{equation}
finally getting 
\begin{equation}
\phi(z)=\frac{\jmath}{\ln{(z+1)+\frac{(1-\jmath)\pi}{2}}}+\frac{\sqrt{\jmath}}{\sqrt{1-\jmath}},\label{pp0}
\end{equation}
immediately, we observe that 
\begin{equation}
z\neq z_{\jmath}=[e^{-\frac{\pi}{2}}]^{1-\jmath}-1\approx (0.20)^{1-\jmath}-1\leq 0, 
\end{equation}
and when $\jmath\to 1$, we obtain $\phi \in (-\mathcal{N},3\mathcal{N}]$, with $\mathcal{N}\gg1$. In \eqref{adfr2}, we have
\begin{align}
F^{-1}_{\phi}&=\frac{(1+\phi^2)^2}{(1+\phi^2)^2-\jmath^{-2}\Big[(1-\jmath)\phi^2-\jmath\Big]\cdot\Big(\phi-\frac{\sqrt{\jmath}}{\sqrt{1-\jmath}}\Big)^4},\label{pp}
\end{align}
and $F^{-1}_{\phi}\neq \{0,\pm \infty\}$ imposing $\jmath \in [\epsilon_1,\epsilon_2]\subset (0,1)$.\footnote{For example, we can set $\epsilon_1=0.3$ and $1-\epsilon_1\geq \epsilon_2$.}

Since two similar worlds present similar characteristic parameters, we emulate the Chevallier-Polarsky-Linder (CPL) model \cite{cpl1,cpl2} which is widely used because the flexibility and robust behavior in describing the evolution of dark energy, leading us to consider, for this sector, an adequate exponential weight function, then for a Moir\'e scenario, we take into account the expression $\ell_\phi=e^{-\vert\phi\vert} $, being able to propose $\Lambda^{h}=\ell_\phi\Lambda^{g}=\bar{\Lambda}$ that satisfies the inherent conditions $\Lambda^{h} \to\{ 0, 1\}\Lambda^{g}$ when $\phi \to\{ \pm\infty, 0\}$, relaxing the tension between the values. 
In addition, the noncommutative proposal \cite{2mg} could be mentioned, where it is possible to show that the $\Lambda$ in a deformed space is proportional to the commutative one via an exponential function $\ell_\eta$, where $\eta$ is the parameter of non-commutativity, and this expression clearly exhibits the contribution of the aforementioned factor. Thus, the confidence region for $\jmath$ is $0\leq\jmath<1$ to avoid singularities in \eqref{pp0}, which is clearly different from the bound imposed in \cite{1mg}.
Now, for the $\Lambda$ associated with the metric $g_{\mu\nu}$, we state $\Lambda^{g}=H^2_0\times 
 \tanh(\phi)$, remarking that the hyperbolic tangent function is strongly supported by the current appearance in emergent DE models \cite{Li_2019, Hernandez-Almada:2020uyr, 3mg}. 
 
Therefore, the dimensionless Friedmann equation becomes the following
\begin{equation} 
   F_{\phi}[E^2(z)-1]=\Omega_{0m}[(z+1)^3-1]+\Omega_{0r}[(z+1)^4-1],\label{adfr}
\end{equation}
where 
\begin{align}
F_{\phi}&=1-\frac{[(1-\jmath)\phi^2-\jmath]}{(1+\phi^2)^2}\Big(\frac{d\phi}{dz}\Big)^2(1+z)^2,
\label{adfr2}
\end{align}
and for $F^{-1}_{\phi}$ with the term \eqref{pp0}, we get \eqref{pp}. 

The deceleration parameter in this model is
\begin{eqnarray}
        q(z)=
        \frac{\Omega_{0m}[2(z+1)^3-1]+\Omega_{0r}[3(z+1)^4-1]-F_ {\phi}}{\big\{\Omega_{0m}[ (z+1)^3-1]+\Omega_{0r}[(z+1)^4-1]+F_ {\phi}\big\}}. \label{DecPar}
\end{eqnarray}
In addition, we write the effective EoS as follows.
\begin{eqnarray}
    &&
    3w_{eff}(z)=2\Big\lbrace\Omega_{0m}[(z+1)^3-1]+\Omega_{0r}[3(z+1)^4-1]\nonumber\\&&-F_ {\phi}\Big\rbrace\big\{\Omega_{0m}[ (z+1)^3-1]+\Omega_{0r}[(z+1)^4-1]+F_ {\phi}\big\}^{-1}\nonumber\\&&-1\,.
\end{eqnarray}
In addition, the parameter space of the model under Moir\'e gravity is $\boldsymbol{\Theta} = \{h,\Omega_{0m},\jmath\}$,
where:
\begin{itemize}
    \item $h \equiv H_0/100$ is the dimensionless Hubble parameter,
    \item     $\Omega_{0m}$ denotes the  
    total matter (including baryons), respectively,
    \item     $\jmath$ is a free parameter of the modified gravity model.
\end{itemize}

%%%%%%%%%%%%%%%%%%%%%%%%%%%%%%%%%%%%%%%%%%%%%%%%%%%%%%
\section{Datasets} \label{sec:constraints}
%%%%%%%%%%%%%%%%%%%%%%%%%%%%%%%%%%%%%%%%%%%%%%%%%%%%%%

We adopt the following priors: a uniform distribution over $h \in [0.2,1]$ and $j \in [0,1]$, plus a Gaussian prior $\Omega_{0m} = 0.3111\pm 0.0056 $ \cite{Planck:2018}. 
The corresponding Gaussian log-likelihood is
\begin{equation}\label{eq:chi2_joint}
    -2\ln(\mathcal{L}_{\rm baseline}) \;\propto\; \chi^2_{\rm baseline} = \sum_i \chi^2_i\,,
\end{equation}
where the sum runs over each statistic $\chi^2$ for the baseline datasets. 
Our analysis employs a Monte Carlo Markov Chain (MCMC) method using the \texttt{Emcee} package \cite{Foreman:2013} in a Python environment, and the chains are required to converge based on the autocorrelation function. The data samples to constrain these parameters are: 
\begin{itemize}
\item The cosmic chronometers (CC) data set comprises 33 uncorrelated $H(z)$ measurements obtained from differential aging of passive galaxies \cite{Moresco:2016mzx, Jiao_2023, Tomasetti_2023}, spanning $0.07<z<1.965$. 
\item Type Ia Supernovae (SNIa). We used the Pantheon+ dataset \cite{Scolnic2018-qf, Brout_2022}, a compilation of 1701 correlated distance modulus measurements on $0.001<z<2.26$.
\item Intermediate-luminosity quasars (QSO). A sample of 120 distance modulus measurements that span a redshift range $0.462<z<2.73$ is presented in \cite{ShuoQSO:2017}.
\item The Hydrogen II Galaxies (HIIG) are compact galaxies with low mass, with the characteristic that their luminosity is dominated by a young massive burst of star formation \cite{Chavez2014}. This allows us to extract a correlation between the luminosity ($L$) and 
the inferred velocity dispersion ($\sigma$) of the ionized gas. A total of 181 distance modulus measurements coming from Hydrogen II galaxies spanning a redshift region $0.01<z<2.6$ are included in the Bayesian analysis \cite{GonzalezMoran2019, Gonzalez-Moran:2021drc}. 
\end{itemize}

%%%%%%%%%%%%%%%%%%%%%%%%%%%%%%%%%%%%%%
\section{Results} \label{Results}
%%%%%%%%%%%%%%%%%%%%%%%%%%%%%%%%%%%%%%

Using the paradigm of Moir\'e gravity, we present the modified Friedmann equation and its coupling to the Moir\'e scalar field via the modified Klein-Gordon(KG) equation, based on the action function. A proposed solution to the KG equation is presented by Eq. \eqref{pp0}, which, together with the Friedmann equation \eqref{adfr}, yields the complete dynamics and evolution of the Universe in this context. We begin our analysis using the parameter space $\boldsymbol{\Theta} = \{h,\Omega_{0m},\jmath\}$ and a region of exploration given by $h\in[0.2,1]$ and $\jmath\in[0,1]$ together with a Gaussian prior for the matter density parameter $\Omega_{0m}$. 

Table \ref{tab:bf_model} summarizes the median values of the free parameters and their $68\%(1\sigma)$ confidence intervals by considering several combinations of CC, SNIa, HIIG, and QSO. The analyzes from the CC + SNIa + QSO (+HIIG) dataset find values of $h$ consistent with the value obtained by Planck \cite{Planck:2018} within $1.4\sigma$ ($1.6\sigma$) while deviations of $1.3\sigma$ ($1.8\sigma$) are achieved for CC + SNIa (+HIIG) samples. Additionally, we observe that the constraint is stronger when a QSO sample is considered in the analysis than when a HIIG sample is included. However, the $1\sigma$ confidence interval for the parameter $\jmath$ is consistent when these samples are included jointly or separately in the CC + SNIa data. The characteristic parameter Moir\'e $\jmath$ is strongly bound to $0.677^{+0.001}_{-0.002}$ when the QSO sample is considered, indicating the presence of a Moir\'e scalar field instead of another kind of scalar field. 

Notice that the value of $\jmath$ indicates the presence of a Moir\'e scalar field instead of a traditional scalar; we expect a non-Moir\'e scalar field if $\jmath=0$. Observe how the scalar field evolves softly, being subdominant while $z\to\infty$. The value of the scalar field has a maximum tendency when $z=0$.

\begin{figure*}
   \centering
   \includegraphics[width=0.22\textwidth]{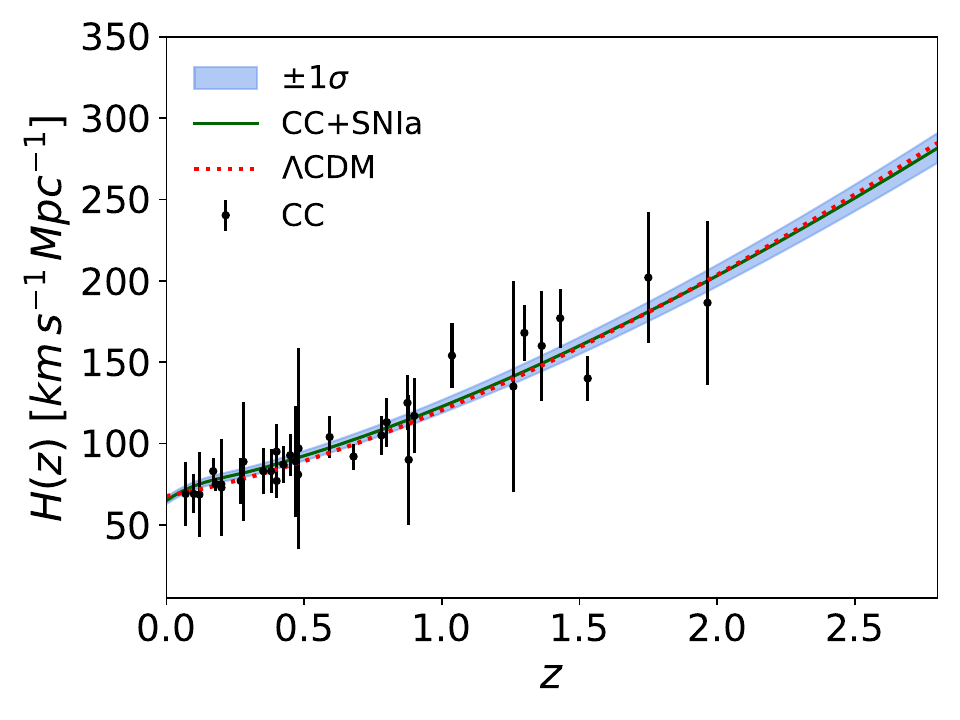}
   \includegraphics[width=0.22\textwidth]{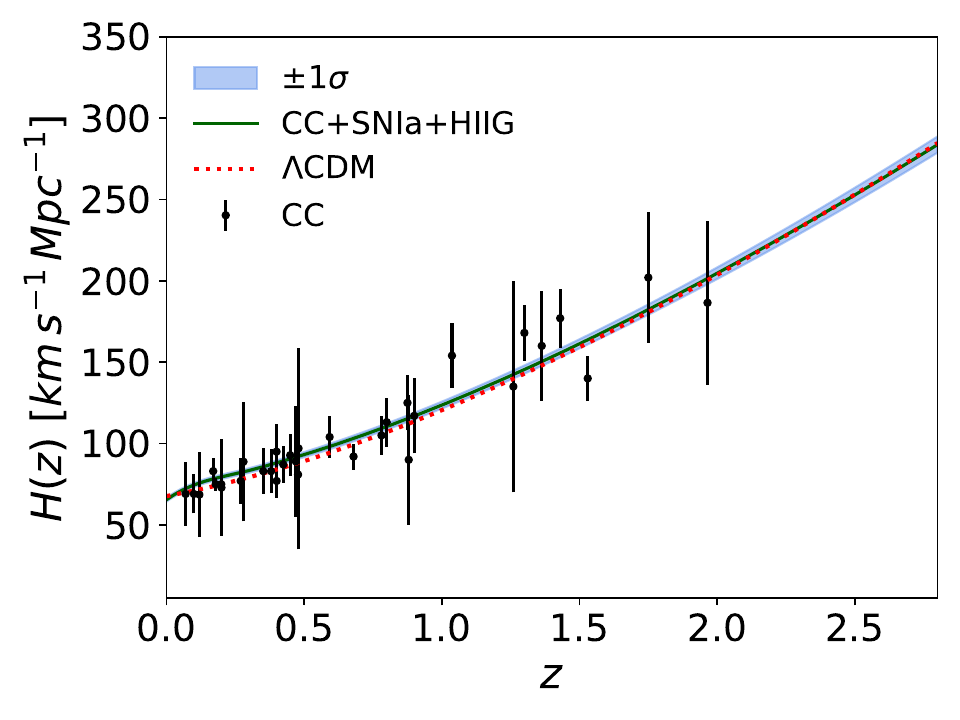}
   \includegraphics[width=0.22\textwidth]{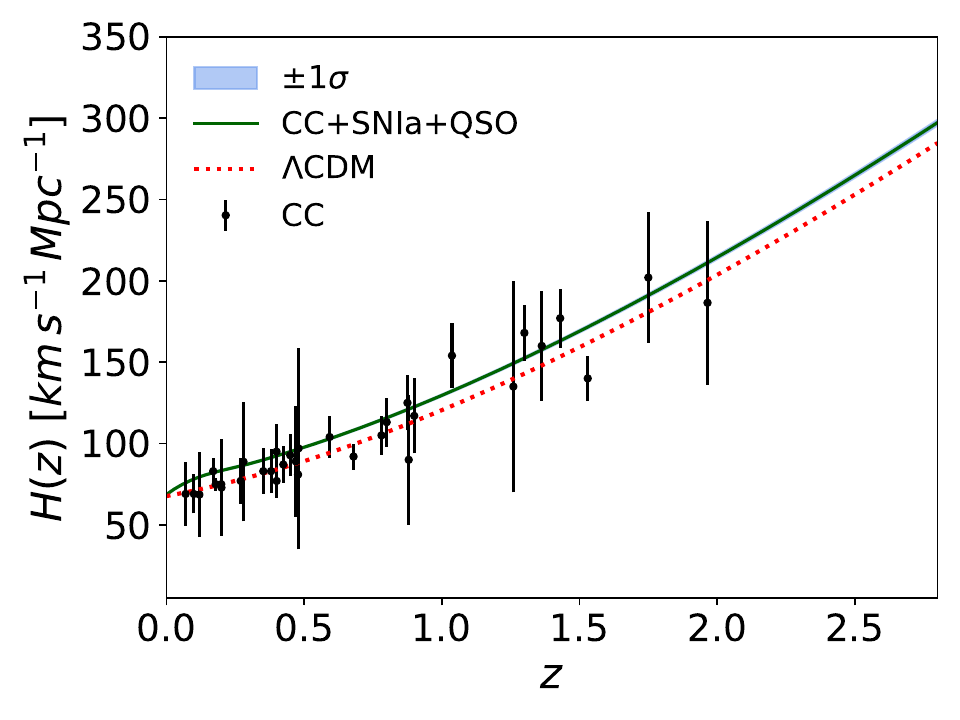}
   \includegraphics[width=0.22\textwidth]{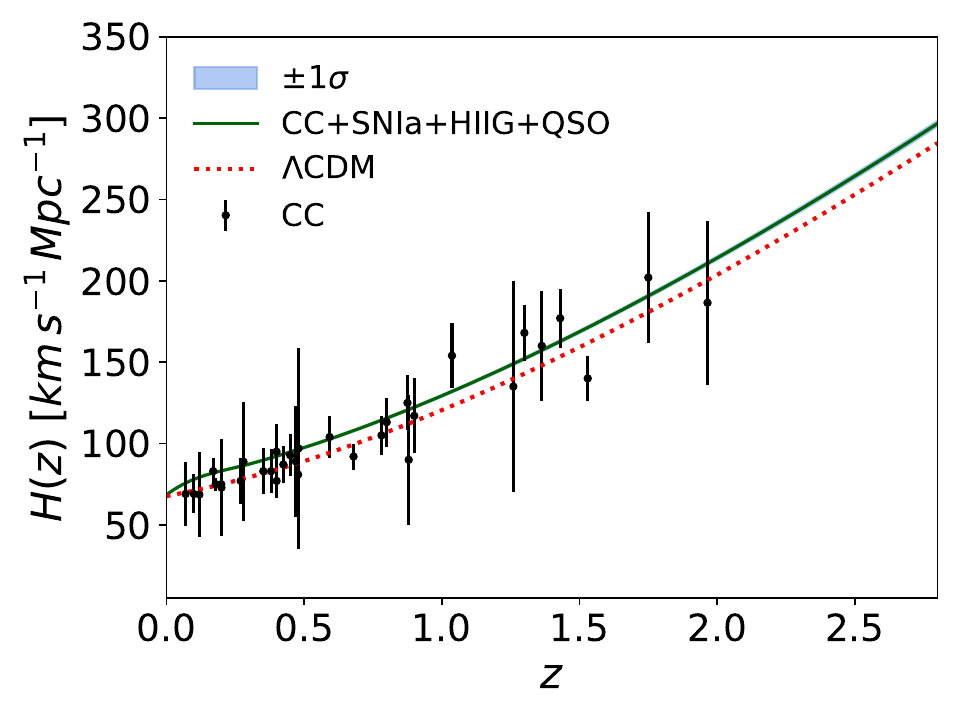}
   \\
   \includegraphics[width=0.22\textwidth]{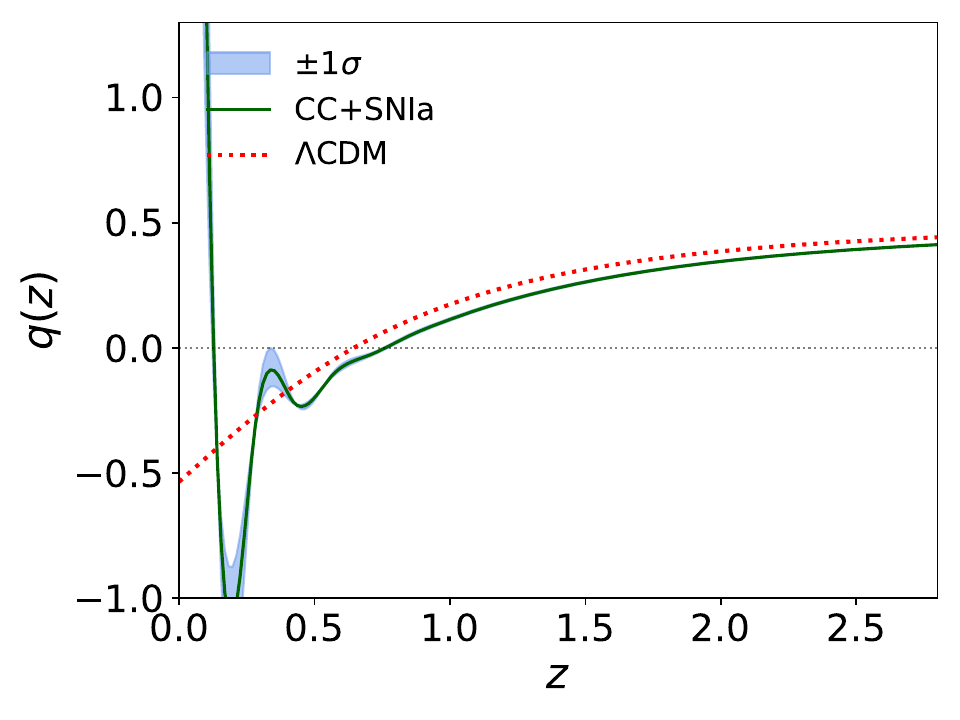}
   \includegraphics[width=0.22\textwidth]{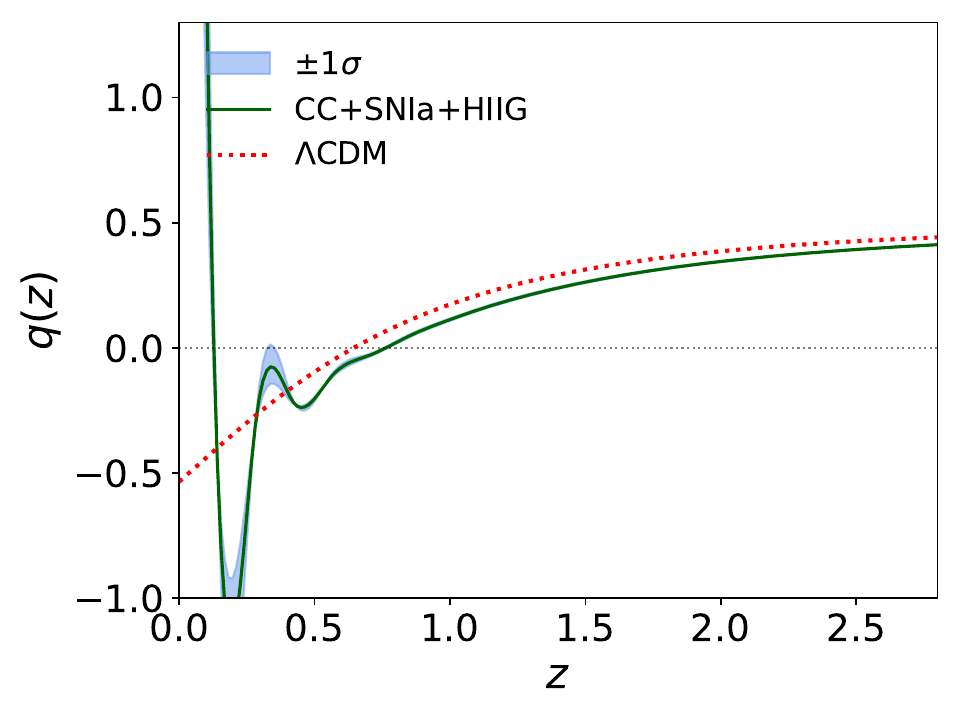}
   \includegraphics[width=0.22\textwidth]{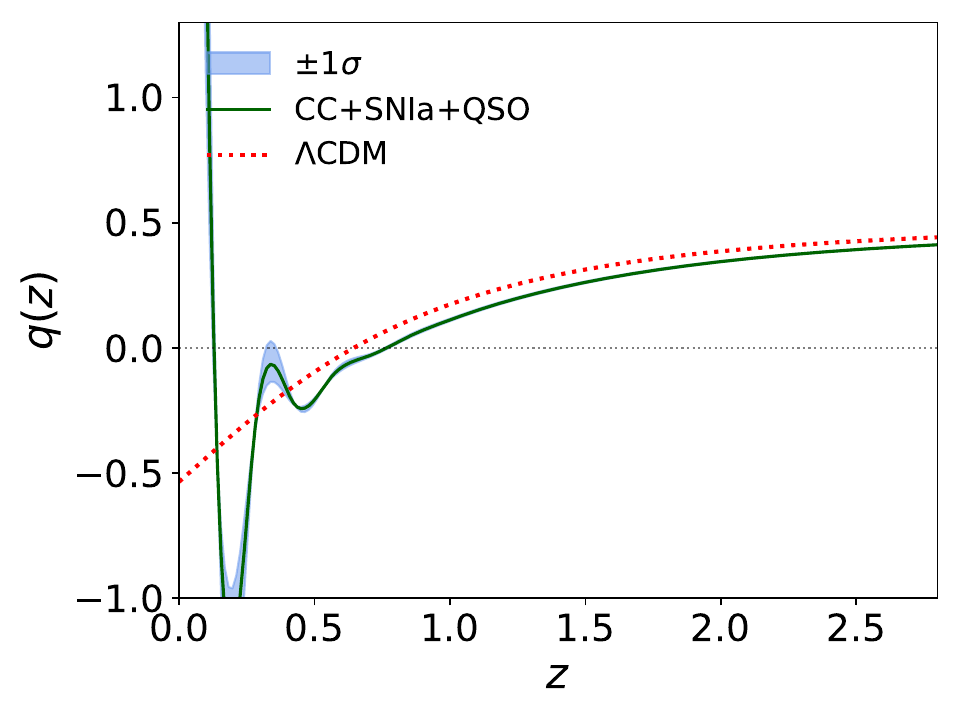}
   \includegraphics[width=0.22\textwidth]{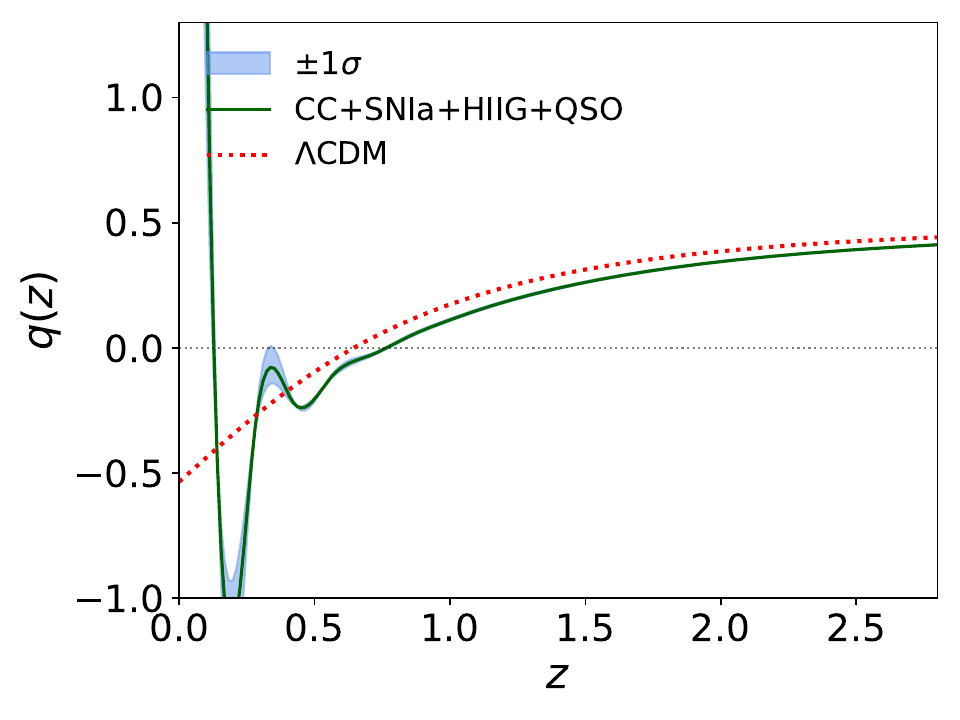}
   \\
   \includegraphics[width=0.22\textwidth]{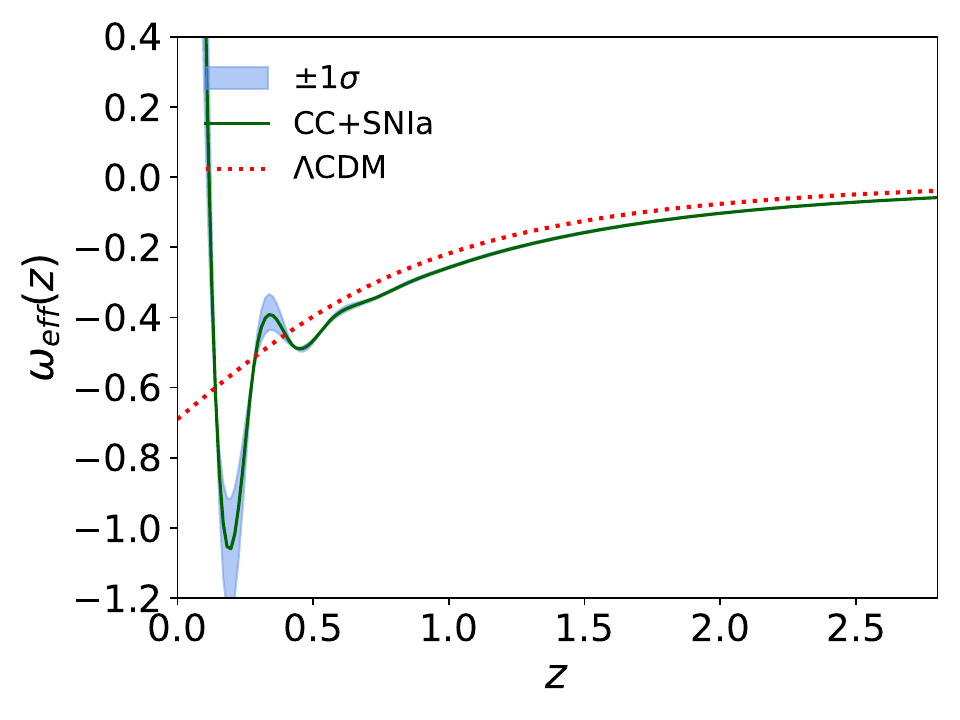}
  \includegraphics[width=0.22\textwidth]{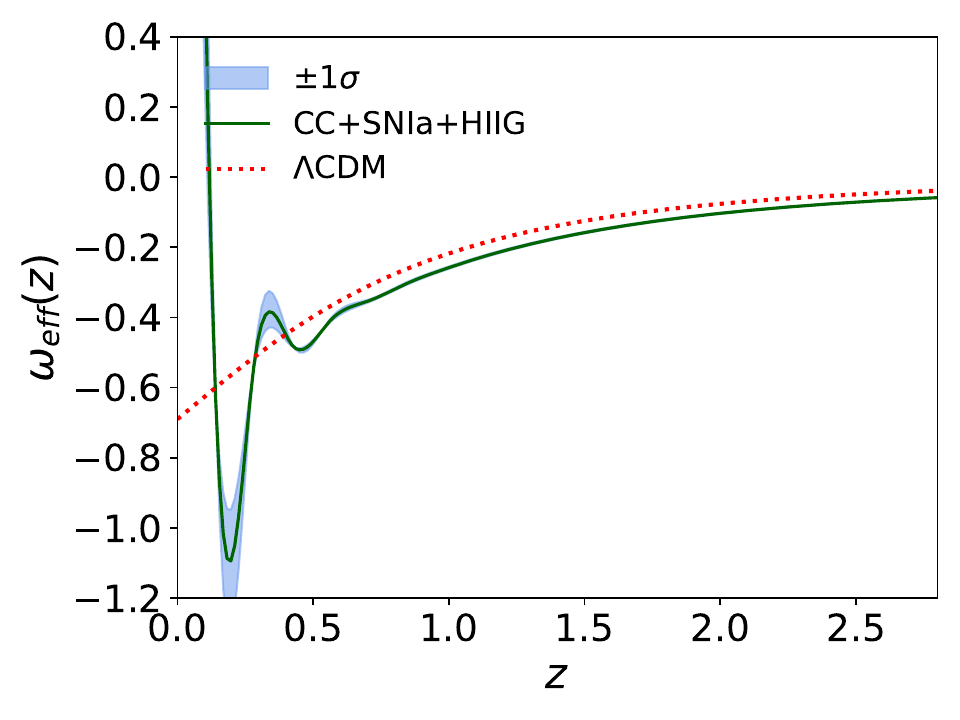}
   \includegraphics[width=0.22\textwidth]{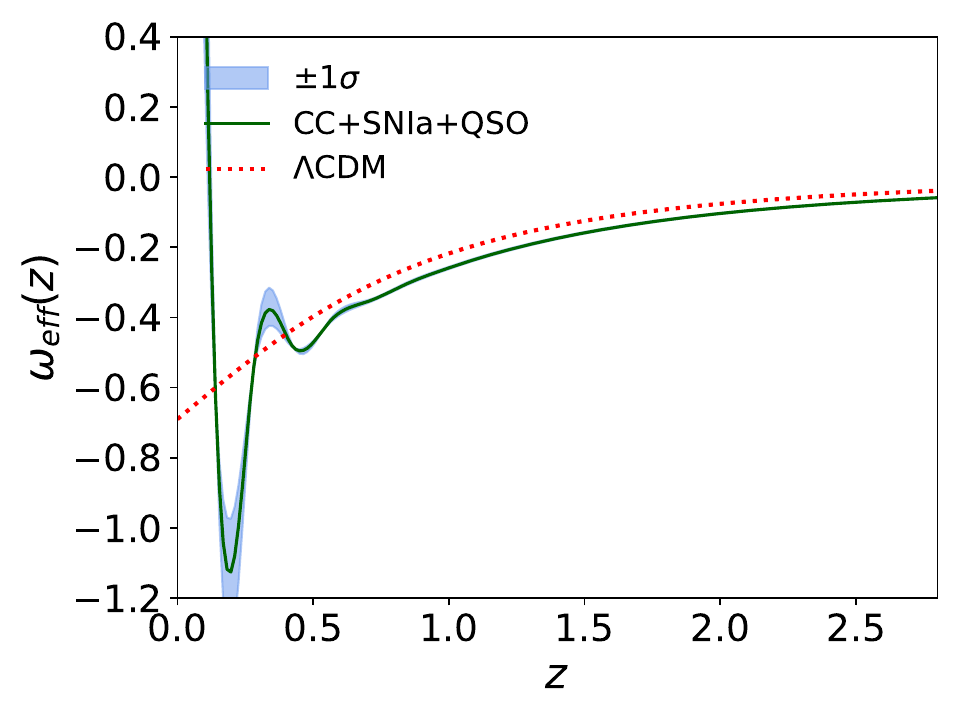}
   \includegraphics[width=0.22\textwidth]{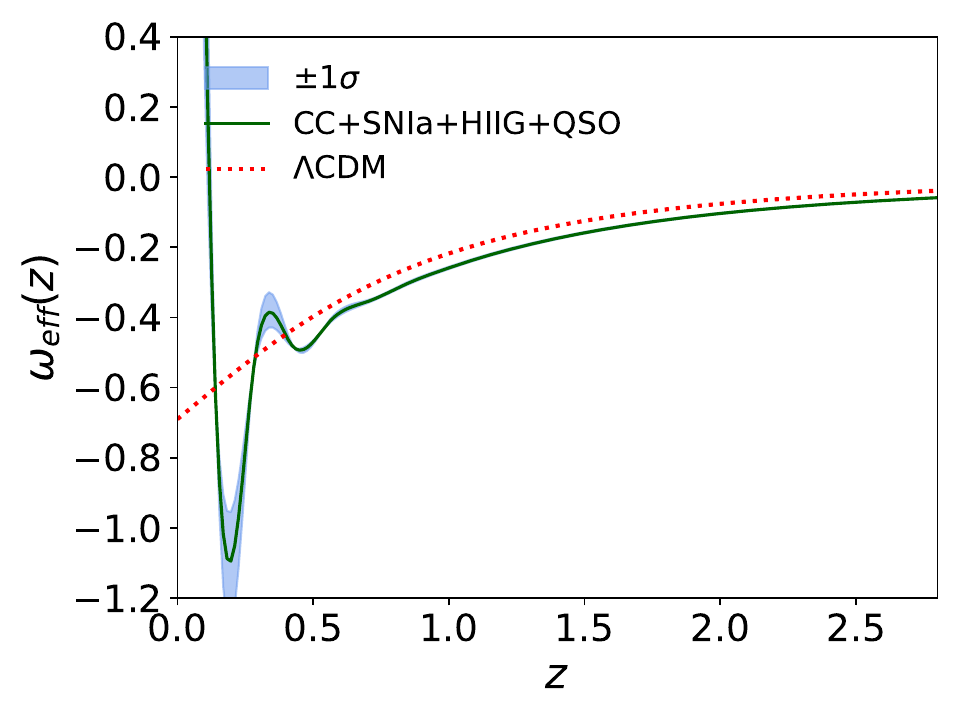}
   \\
   \includegraphics[width=0.22\textwidth]{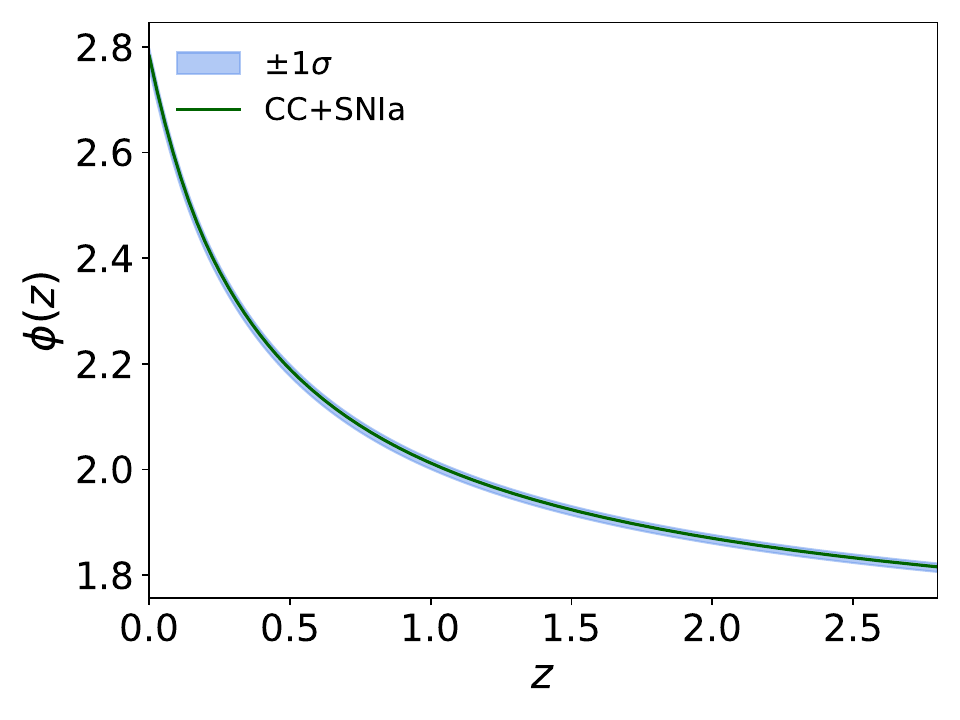}
   \includegraphics[width=0.22\textwidth]{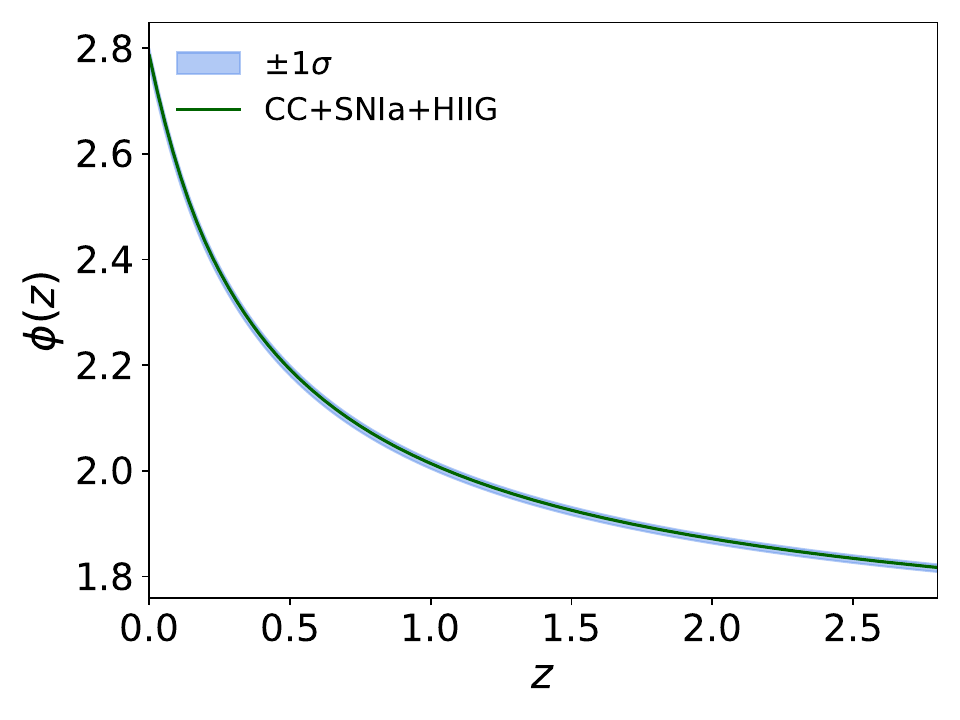}
   \includegraphics[width=0.22\textwidth]{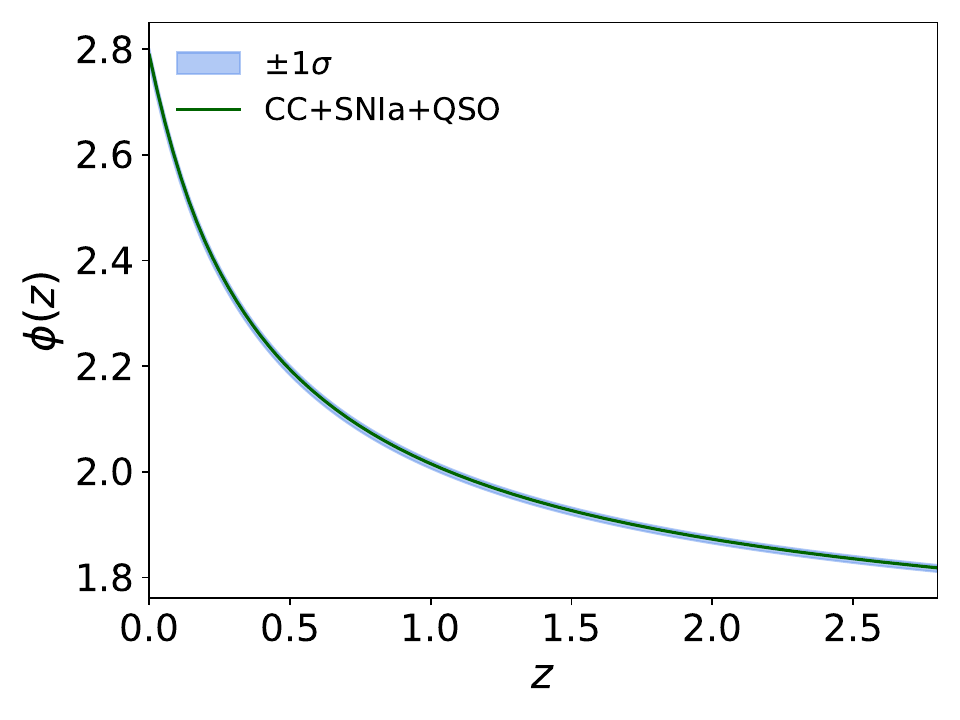}
   \includegraphics[width=0.22\textwidth]{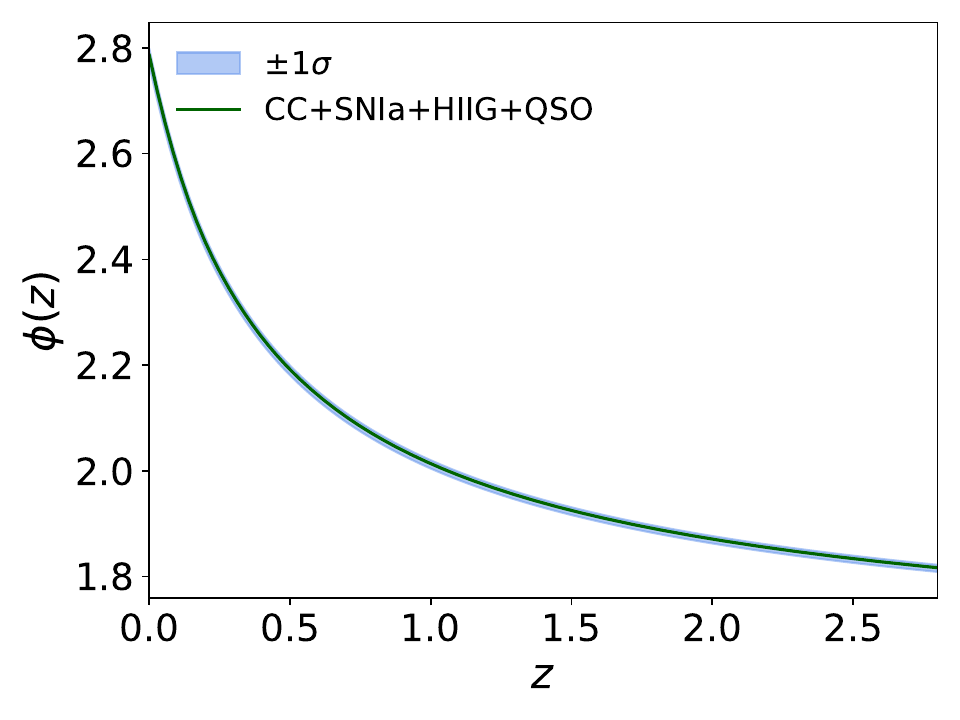}
   \caption{Reconstructions of the expansion history for our model parametrization  using four data combinations (left to right): CC+SNIa, CC+SNIa+HIIG, CC+SNIa+QSO and CC+SNIa+HIIG+QSO. From top to bottom, we show the Hubble parameter $H(z)$, the deceleration parameter $q(z)$, the effective EoS $w_{eff}(z)$ and the Moir\'e scalar field $\phi(z)$. Solid curves correspond to the posterior median reconstruction, with shaded regions indicating the $1\sigma$  intervals. The reference $\Lambda$CDM prediction is overplotted as a red dashed curve.}
   \label{fig:cosmography}
\end{figure*}

Furthermore, we find a redshift of the deceleration-acceleration transition $z_T=0.759\pm0.011$ using the full data (joint)  and in agreement within $1\sigma$ with the other data combinations. This result suggests a transition earlier than that expected by $\Lambda$CDM ($z_T=0.630\pm 0.015$ \cite{Astorga-Moreno2025-oz}) that deviates in  $6.9\sigma$. In contrast to $\Lambda$CDM, in the Moir\'e gravity framework, there are two deceleration phases, as shown in Figure \ref{fig:cosmography}, which present an extra abrupt transition to the deceleration region around $z\sim 0.1$. As a consequence, the current value of the deceleration parameter $q_0$ is positive $q_0 \sim 15.5$. This could be caused by the Moir\'e scalar field increasing its value as $z\to 0$ (see Fig. \ref{fig:cosmography}). 
Moreover, observing the behavior of $w_{eff}$ in the final stages of the Universe's evolution, we find a dynamical dark energy that transitions between quintessence, a cosmological constant, and even a phantom region. Notice that the evolution of $w_{eff}$ rapidly transitions to a gravitationally attractive fluid, producing the deceleration observed in the $q(z)$ parameter. 

Additionally, Table \ref{tab:bf_model} shows the age of the Universe ($\tau$), finding a deviation from the value obtained by CMB Planck \cite{Planck:2018} within $2\sigma$  for CC + SNIa (+HIIG). However, when the QSO sample is added to these datasets, the cosmology of Moir\'e suggests a $\tau_U \sim 12.9\,$Gyrs, which is much lower than expected by CMB Planck based on the $\Lambda$CDM model. Nevertheless, notice something important. In Table \ref{tab:bf_model}, the joint analysis shows an $H_0$ value compatible with Planck results, despite the fact that we use a uniform prior. Suggesting an alleviation of the Hubble tension. For example, using supernova data samples by Riess \cite{Riess_2022} gives an $H_0$ value incompatible with Planck and, according to the $\Lambda$CDM model, a younger Universe in disagreement with globular cluster observations \cite{OldestStar}. However, a remarkable point is that the Moir\'e model outperforms $\Lambda$CDM when used with SNIa data, as it alleviates the Hubble tension, even though it implies a younger Universe, which is not consistent with the standard accepted age of the Universe. As a final point, note also that recent studies suggest a universe age of $26.7$ Gyr \cite{deAndres:2024vmr}, almost twice that predicted by the standard cosmological model, generating an important tension across almost all alternatives to $\Lambda$CDM.

\begin{table*}[ht!]
	\centering
	\caption{Median values and their $1\sigma$ confidence interval. Here it is shown the dimensionless Hubble constant, matter density parameter, $\jmath$ parameter, the age of the universe, the transition redshift and the deceleration parameter at $z=0$.}
	\label{tab:bf_model}
	\begin{tabular}{lccccccc} 
    \hline
    Data & $\chi^2$ & $h$ & $\Omega_{0m}$ & $\jmath$ & $\tau_U\,$[Gyrs]    & $z_T $ & $q_0$   \\ [0.9ex] 
    \hline
    \multicolumn{8}{c}{} \\ [0.9ex]
CC+SNIa &1976.49 & $0.649^{+0.021}_{-0.021}$  & $0.314^{+0.005}_{-0.005}$  & $0.677^{+0.002}_{-0.002}$  & $13.560^{+0.448}_{-0.426}$  & $0.755^{+0.012}_{-0.011}$  & $14.331^{+6.546}_{-4.768}$  \\ [0.9ex] 
CC+SNIa+HIIG &2413.15 & $0.654^{+0.012}_{-0.011}$  & $0.314^{+0.005}_{-0.005}$  & $0.677^{+0.001}_{-0.002}$  & $13.453^{+0.231}_{-0.226}$  & $0.757^{+0.012}_{-0.011}$  & $15.402^{+6.717}_{-4.897}$  \\ [0.9ex] 
CC+SNIa+QSO &5104.96 & $0.686^{+0.004}_{-0.004}$  & $0.313^{+0.005}_{-0.005}$  & $0.678^{+0.001}_{-0.002}$  & $12.831^{+0.058}_{-0.058}$  & $0.759^{+0.012}_{-0.012}$  & $16.353^{+7.056}_{-5.077}$  \\ [0.9ex] 
CC+SNIa+HIIG+QSO &5546.77 & $0.685^{+0.004}_{-0.004}$  & $0.313^{+0.005}_{-0.005}$  & $0.677^{+0.001}_{-0.002}$  & $12.857^{+0.057}_{-0.057}$  & $0.759^{+0.011}_{-0.011}$  & $15.376^{+6.403}_{-4.692}$  \\ [0.9ex]
\hline
	\end{tabular}
\end{table*}

\begin{figure*}
   \centering
   \includegraphics[width=0.6\textwidth]{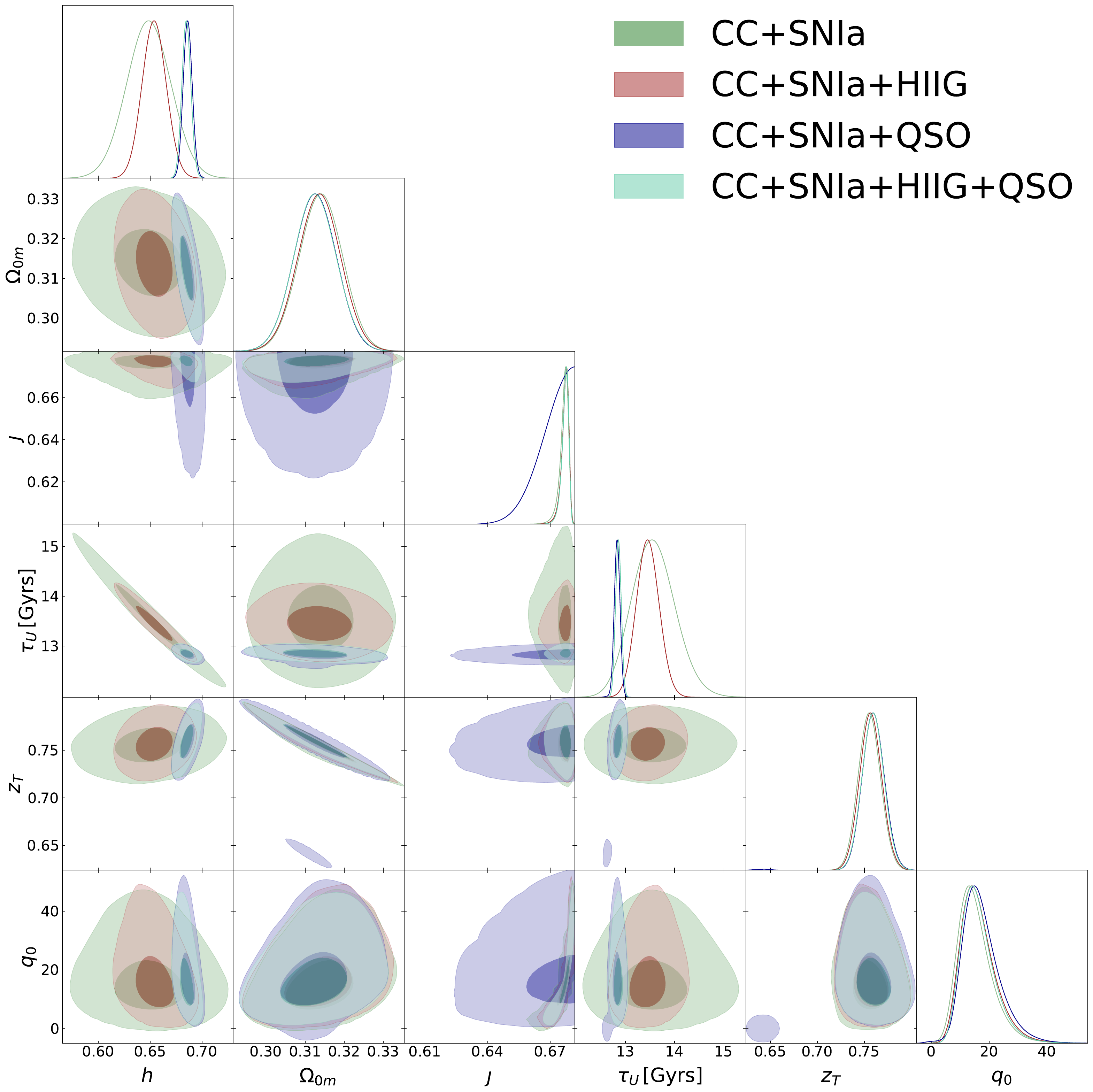}
   \caption{Posterior distributions for the Moir\'e gravity parameters $\boldsymbol{\Theta }= (h, \Omega_{0}, \jmath)$ obtained from the MCMC analysis. Diagonal panels show the marginalized 1D posteriors, while off-diagonal panels display the joint 2D regions at $1\sigma$ and $2\sigma$ confidence levels.}
 \label{fig:contours}
\end{figure*}

%%%%%%%%%%%%%%%%%%%%%%%%%%%%%%%%%%%%%%
\section{Conclusions and Discussions} \label{SD}
%%%%%%%%%%%%%%%%%%%%%%%%%%%%%%%%%%%%%%
Remarkable ideas surge from solid state physics, and its application to cosmology is not the exception. In this case, Moir\'e gravity arises from ideas of solid-state layers, with an interesting correspondence to brane-world theories \cite{Maartens:2010ar}. Naturally, topological configurations like these, related to Moir\'e gravity, allow the existence of a scalar field, called in this case the Moire scalar field, maintaining similarity to the radion scalar field in theories of extra dimensions \cite{Maartens:2010ar}
The goal is to establish the theoretical relationship between graphene-inspired approaches to emergent gravity \cite{new5,new6} and to explore how large-scale cosmological observations can be used to constrain the microscopic quantum state underlying the emergence of spacetime.
Additionally, the Friedmann equation presents modifications caused by this new topological configuration coupled with the Moir\'e scalar field.
In this vein, we show an analysis of the Moir\'e gravity applied to cosmology, in order to constrain the free parameters of the Moir\'e scalar field and compare with those previously reported in the literature. As a starting point, we integrate the Klein-Gordon equation to obtain a plausible solution for the Moir\'e scalar field, avoiding singularities for $z\geqslant0$.  
The modified Friedmann equation is contrasted with recent observations of cosmic chronometers, supernovae of Ia type, intermediate-luminosity quasars, and hydrogen II galaxies, summarizing the results in Table \ref{tab:bf_model} where for a joint analysis the free parameter of the Moir\'e scalar field takes the medium value $j=0.677$ in contrast to that reported in the literature \cite{1mg} having a bound of $-\infty<j<\frac{1}{2}$. From our analysis, we observe interesting features in the joint case. First, the age of the Universe is lower than expected by the $\Lambda$CDM model, raising concerns mainly because recent observations detect galaxies older than the Universe, even in $\Lambda$CDM. This behavior is understandable when we observe the parameter $q_0$ in Table \ref{tab:bf_model}, having a medium value of $15.376$ at $z=0$, implying a decelerating Universe. However, notice that this value is extremely high compared with other models that study the decelerating dark energy near $z=0$. As we mentioned previously, there is $\sim3.4\sigma$ of consistency compared with the $\Lambda$CDM model. Additionally, a decelerating universe in its last stages has a better fit with DESI than models that never decelerate, like the $\Lambda$CDM model (see, for instance, \cite{DESI:2025zgx}). In addition, we need to keep in mind that Eq. \eqref{pp0} is just an appropriate approximation (in order to avoid singularities) instead of a full solution. Thus, a deeper analysis is suggested in future studies. Notice also that we have two peaks in the acceleration, producing an important disturbance in the space-time evolution. Nevertheless, we refer to Fig. \ref{fig:cosmography} for $w_{eff}$ and deduce how the dark energy evolves from quintessence, passes into a cosmological constant, and ultimately enters the phantom regime. After that, the evolution rapidly degenerates into an attractive fluid that suddenly decelerates the Universe, as is also observed by $q(z)$. The singularity that contains the Moir\'e scalar field equation could pose problems for deeper analysis using perturbation theory. For example, it could generate a ghost field or other pathologies. Even the extreme accelerated and decelerated phase, together with the Moir\'e scalar field, could generate other pathologies that can affect the general structure of the theory. Nevertheless, if we want to comprehend whether the theory is free of ghosts, tachyons, or others, it is necessary a perturbative/quantum analysis. However, this is outside of the scope of this paper. In this vein, future low-redshift observations and a perturbative analysis of the Moiré sector will be necessary to determine whether this behavior is physically viable or is instead a consequence of the limited constraining power of the present background data.
Finally, we also remark from the reconstructions given in Fig. \ref{fig:cosmography}, focusing on the behavior of the Moir\'e field, that emulates the expansion of the universe, since as $z$ is big enough we find that $\phi\ll 1$ or is reaching the singularity (Big Bang epoch), and as has been stated in Ref. \cite{1mg}, yields the expected history for the universe, starting with an inflationary period and reaching a small vacuum energy density that continuously tends to zero.

%%%%%%%%%%%%%%%%%%%%%%%%%%%%%%%%%%%%%%
\begin{acknowledgments}
We thank the anonymous referee for thoughtful remarks and suggestions. J.A.A.-M. acknowledges SECIHTI for support by a postdoctoral fellowship at Cinvestav, M\'exico. M.A.G.-A. acknowledges support from c\'atedra Marcos Moshinsky, SECIHTI for the support with the National Research System (SNII) grant and the project 0056 from Universidad Iberoamericana: Nuestro Universo en Aceleraci\'on, energ\'ia oscura o modificaciones a la relatividad general. The numerical analysis was also carried out by {\it Numerical Integration for Cosmological Theory and Experiments in High-energy Astrophysics} (Nicte Ha) cluster at IBERO University, acquired through c\'atedra MM support.  A.H.A. thanks to the support from Luis Aguilar, 
Alejandro de Le\'on, Carlos Flores, and Jair Garc\'ia of the Laboratorio 
Nacional de Visualizaci\'on Cient\'ifica Avanzada. A.H.A and M.A.G.-A acknowledge partial support from project ANID Vinculaci\'on Internacional FOVI220144. 
\end{acknowledgments}
\bibliography{main}
\end{document}